\documentclass[12pt,preprint]{aastex}

\def\t0{\theta_{\circ}}

\def\be{\begin{equation}}
\def\en{\end{equation}}

\begin{document}

\shorttitle{New sub-mm observations of 55 Cancri}
\shortauthors{Jayawardhana et al.}

\title {New sub-millimeter limits on dust in the 55 Cancri planetary system}
\author{Ray Jayawardhana\altaffilmark{1},
Wayne S. Holland\altaffilmark{2},
Paul Kalas\altaffilmark{1},
Jane S. Greaves\altaffilmark{2},\\
William R. F. Dent\altaffilmark{2},
Mark C. Wyatt\altaffilmark{2}, 
and Geoffrey W. Marcy\altaffilmark{1}}
\altaffiltext{1}{Department of Astronomy, University of California, Berkeley, CA 94720, U.S.A.}
\altaffiltext{2}{UK Astronomy Technology Centre, Royal Observatory, Blackford Hill, Edinburgh EH9 3HJ, United Kingdom}

\begin{abstract}
We present new, high-sensitivity sub-millimeter observations towards 55 
Cancri, a nearby G8 star with one, or possibly two, known planetary 
companion(s). 
Our 850 $\mu$m map, obtained with the SCUBA instrument on the 
James Clerk Maxwell Telescope, shows three peaks of emission at the 2.5 mJy 
level in the vicinity of the star's position. 
However, the observed peaks are 25$\arcsec$--40$\arcsec$ away from the star 
and a deep $R$-band optical image reveals faint point sources
that coincide with two of the sub-millimeter peaks. Thus, we do not find  
evidence for dust emission spatially associated with 55 Cancri. The 
excess 60 $\mu$m emission detected with ISO
may originate from one
or more of the 850 $\mu$m peaks that we attribute to background sources.
Our new results, together with the HST/NICMOS 
coronographic images in the near-infrared, place stringent limits on the 
amount of dust in this planetary system, and argue against the
existence of a detectable circumstellar
dust disk around 55 Cnc.
\end{abstract}


\section{Introduction}
Dusty disks that are believed to be the debris of planetary formation 
have now been imaged at infrared and sub-millimeter wavelengths
around several nearby main-sequence stars including $\beta$ Pictoris,
HR 4796A, Vega, Fomalhaut and $\epsilon$ Eridani (Smith \& Terrile 1984; 
Holland et al. 1998; Jayawardhana et al. 1998; Greaves et al. 1998). 
These disks may be young analogs of the Kuiper
Belt in our solar system (Jewitt \& Luu 2000). The $\sim$5-Gyr-old (Gonzalez \& Vanture 1998; 
Baliunas et al. 1997) G8V star 55 Cancri (HR 3522; HD 75732), at a distance 
of 13 pc, has attracted much attention in recent years for having planet(s) 
as well as a possible dust disk. The only other example reported to date of 
a planet co-existing with a disk is $\epsilon$ Eri, which has a well-resolved, 
prominent disk in the sub-millimeter images (Greaves et al. 1998). 
However, the presence of a giant planet around $\epsilon$ Eri (Hatzes 
et al. 2000) is difficult to establish with radial velocity measurements 
because of the high level of stellar chromospheric activity and possible 
face-on orientation of the planet orbit. On the other hand, in the case 
of 55 Cnc, it is the putative disk that is in dispute.

55 Cnc harbors an inner planet with {\it M sin i} = 0.84 M$_{Jup}$ in an 
orbit with a semi-major axis of 0.11 AU (Butler et al. 1997). There is 
evidence for a second planet at several AU in the form of a residual drift 
in the radial velocity over the past 10 years (Marcy \& Butler 1998; Fischer 
et al. 2001). It has been noted that 55 Cnc --along with 14 Her, another 
planet-bearing star-- is one of the two most metal-rich stars in the solar 
neighborhood, with [Fe/H]$\ge$+0.4, perhaps suggesting a causal link between 
planets and the metallicity of the parent star (Gonzalez \& Vanture 1998; 
Gonzalez, Wallerstein \& Saar 1999).

Dominik et al. (1998) first presented evidence for a Vega-like disk around 
55 Cnc based on {\it Infrared Space Observatory} (ISO) measurements between 
25 $\mu$m and 180 $\mu$m. They detected the stellar photosphere at 25 $\mu$m,
significant excess emission at 60 $\mu$m, and infrared cirrus at 135 and 180 $\mu$m.
Trilling \& Brown (1998) and Trilling, Brown \& 
Rivkin (2000) confirmed the debris disk interpretation of the far-infrared
data by reporting a resolved
scattered light disk out to a radius of 3.24$\arcsec$ (40 AU) in ground-based, 
near-infrared, coronographic images.  They estimated the disk position
angle as 50 $\pm$ 10 degrees, inclination 27$^{+8}_{-11}$ degrees and
minimum dust mass 0.4 M$_{Earth}$. 
Our previous sub-millimeter observations detected 850 $\mu$m flux 
of 2.8$\pm$0.5 mJy and 450 $\mu$m flux of 7.9$\pm$4.2 mJy in the direction of 
55 Cnc, which implied a dust mass at least 100 times lower than the 
Trilling \& Brown (1998) estimate. Our mid-infrared observations at 
10 $\mu$m and 18 $\mu$m showed no excess emission above the stellar 
photospheric 
level (Jayawardhana et al. 2000). Using the NICMOS near-infrared camera
on the {\it Hubble Space Telescope} (HST), Schneider et al. (2001) were 
not able to confirm the results of Trilling and co-workers down to a flux 
level that is 10 times lower.

Here we report new, extremely sensitive sub-millimeter observations in the 
vicinity of 55 Cnc using the Submillimeter Common User Bolometer Array 
(SCUBA) on the James Clerk Maxwell Telescope (JCMT). We find three faint
sources of emission, none of which is centered exactly on 55 Cnc. Our results, 
combined with the NICMOS observations, place strong limits on the amount
of dust emission associated with the 55 Cnc planetary system.

\section{Observations and Results}
The 850 $\mu$m observations of 55 Cnc presented here were carried out with the 
SCUBA instrument (Holland et al. 1999) on the JCMT on Mauna Kea, Hawaii. The 
data were obtained over several observing runs in 1999-2000 using the SCUBA 
photometry mode. In total, there were 31 separate photometric observations 
at two different pointings in the vicinity of 55 Cnc for a total on-source 
integration time of 7.1 hours. Zenith atmospheric opacities at 850 $\mu$m 
ranged from 0.10 to 0.35. When coadded together, the maps were noise
weighted to account for the range of conditions.
Observations of Uranus were 
used for calibrations. Pointing accuracy was 2$\arcsec$, which is small 
compared with the beam size of 15$\arcsec$ at 850 $\mu$m 
(full-width at half-maximum; FWHM). The data 
were reduced using the SCUBA User Reduction Facility (Jenness \& Lightfoot 
1998). The co-added map is shown in Figure 1.

The 850 $\mu$m map shows three obvious peaks of emission at the $\sim$2.5 mJy
level. Their fluxes and positions are given in Table 1. The rest of the field
is rather flat with an rms noise of about 0.4 mJy. Each of the peaks is 
25$\arcsec$--40$\arcsec$ (325--500 AU) away from 55 Cnc. There is no emission 
exactly coincident with the position of 55 Cnc. 

We also obtained optical $R$-band CCD images at the University of Hawaii 
2.2-meter telescope on Mauna Kea on 31 January 2000. An optical coronograph
(Kalas \& Jewitt 1996) blocked the stellar point-spread-function (PSF) at 
the focal plane using a hard-edged occulting spot with 6.5$\arcsec$ diameter.  
A Lyot stop at a re-imaged pupil plane suppressed diffracted light from the 
telescope mirrors and support structures.  The imaging camera was a TEK 
2048$\times$2048 CCD with a plate scale of 0.407$\arcsec$ per pixel and 
with a circular field of view 5.5$\arcmin$ in diameter. 
Thirty-two frames with 20 seconds integration
time each were obtained for 55 Cnc.  Immediately after these observations we
imaged a nearby bright star HR 3521 with the same instrumental configuration.
These data serve as a reference for subtracting the 55 Cnc PSF.

After bias-subtraction and flatfielding, we selected 20 frames from the 55 Cnc
data that were well-centered behind the occulting spot and that had field star
PSF's with FWHM in the range 1.0$-$1.2$\arcsec$.  These 
frames
were median-combined to produce an image of 55 Cnc with 400 seconds cumulative 
integration time.  The HR 3521 data was for the most part unsuitable for 
subtracting the 55 Cnc PSF due to poor centering of HR 3521 behind the 
occulting spot. However, one HR 3521 frame (15 seconds integration) was 
well-centered and adopted as the PSF reference source.  We azimuthally 
smoothed this PSF by rotating the image in 20 degree increments and 
median-combining the resulting frames.  The azimuthally smoothed PSF was 
subtracted from the original HR 3521 frame, revealing residual light
along a southeast-northwest axis.  The residual halo of the self-subtracted 
PSF is probably due to a combination of instrumental and atmospheric 
aberrations.  

The azimuthally smoothed HR 3521 PSF was registered, scaled and subtracted 
from 55 Cnc (Fig. 2).  The subtracted 55 Cnc PSF also appears to have 
residual light symmetrically distributed about the occulting spot along a 
southeast-northwest axis.  This axis is perpendicular to the position angle 
of the disk reported by Trilling \& Brown (1998). The residual PSF halo 
extends from approximately 4$\arcsec$ to 16$\arcsec$ radius, decreasing
approximately with the fourth power of radius, and with surface brightness 
in the range 17.5$-$23.0 mag arcsec$^{-2}$. Trilling \& Brown (1998) report 
an H-band disk surface brightness that falls below 24 mag arcsec$^{-2}$ 
beyond 4$\arcsec$ radius. Our data do not have the sensitivity to detect 
extended nebulosity fainter than 23.0 mag arcsec$^{-2}$.  

We performed a second PSF subtraction using an azimuthally smoothed version 
of the 55 Cnc image as described above for HR 3521.  This self-subtraction 
of 55 Cnc also resulted in residual light along a southeast-northwest axis.  
Because the residual light appears in both the 55 Cnc and HR 3521 
self-subtractions, we attribute it purely to instrumental and atmospheric 
aberrations.  The dominant source is most likely a static aberration such 
as focus error. Finally, as a third PSF subtraction, we selected the best 
55 Cnc frame and registered, scaled and subtracted the unsmoothed HR 3521 
frame.  Though the result contains significantly greater noise than the 
image in Fig. 2, the residual light does not contain significant azimuthal 
asymmetry.  

Three point sources are detected in the $R$-band image within a 40$\arcsec$ 
radius surrounding 55 Cnc (Fig. 2). Their positions are indicated by 
white boxes in Fig. 1 to show their relation to the 850 $\mu$m emission. 
Table 1 lists the magnitudes of the $R$-band sources and their offsets in 
arcseconds from the centroids of 850 $\mu$m emission. The two point sources 
north of 55 Cnc may be optical counterparts of the 850 $\mu$m sources. 
No object was detected in the optical data that corresponds to the position 
of the sub-millimeter peak south of 55 Cnc. The 3$\sigma$ point source 
sensitivity of the optical data is m$_R$ = 22.2 mag, as determined by
injecting artificial point sources near the positions of the optically 
detected sources and the 850 $\mu$m peaks. 

\section{Discussion}
Our sub-millimeter observations of the field around 55 Cnc reveal three 
faint sources of emission at the $\sim$2.5 mJy level. The separation 
between 55 Cnc and each of the 850 $\mu$m peaks is approximately two beam 
widths and thus unambiguously non-coincident with the star.
The spatial association of two of the sub-millimeter peaks with optically
detected point sources (Fig. 2) makes it unlikely that the 
850 $\mu$m peaks are components of an extended dust ring around 55 Cnc.

According to the galaxy counts derived by Blain et al. (1999) from their
deep sub-millimeter survey, one would expect 3$\pm$1 galaxies brighter
than 1 mJy at 850 $\mu$m in a patch of the sky with a 40$\arcsec$ radius. 
Thus, all of the peaks in our 850 $\mu$m map could be background galaxies.
In that case, the optical point sources could either be bright compact
nuclei of the galaxies or stars whose positional coincidence with the 
sub-millimter peaks is simply due to chance alignments in the line of sight.
Without additional color information or spectra, 
we are not able to comment further on the nature of these sources. 

If the sub-millimeter peaks are due to unrelated sources, which appears 
likely, then we can place an upper limit of $\sim$0.4 mJy to the 850 $\mu$m 
emission from within $\sim$100 AU of 55 Cnc itself. Since the sub-millimeter 
flux is relatively insensitive to the temperature of dust grains, we can use 
it to derive a limit on the dust mass associated with 55 Cnc. Following Jura 
et al. (1995), the dust mass $M_d$ is given by

\begin{equation}
M_d = F_{\nu} R^2 \lambda^2 / \left [2kT_{gr} K_{abs} \left (\lambda \right )\right ],
\end{equation}
if $R$ denotes the distance from the sun to 55 Cnc.  Assuming a dust 
absorption coefficient $K_{abs}(\lambda)$ between 1.7 and 0.4 cm$^2$ g$^{-1}$
at 850 $\mu$m (Greaves et al. 1998), we obtain an upper limit to the
dust mass of 1$-$7$\times$10$^{-4}$ M$_{Earth}$,
assuming a disk temperature range $T$= 100-130 K. The lower value of 
$K_{abs}(\lambda)$ is suggested by models of large, icy grains (Pollack 
et al. 1994), while the higher estimate has been used for previous 
observations of debris disks (Holland et al. 1998). It is important to 
point out that sub-millimeter observations are not sensitive to very large 
grains or planetesimals, which could dominate the total mass while adding 
little sub-millimeter emission. Therefore, our dust mass estimate is only 
a lower limit to the total mass, and does not rule out a Kuiper Belt analog 
around 55 Cnc comprised of larger particles.

HST/NICMOS coronographic observations by Schneider et al. (2001) at 1.1 $\mu$m
did not detect a dust disk around 55 Cnc down to a flux level 10 times below 
that reported by Trilling and co-workers. Schneider et al. also searched for 
and failed to find a suggested flux-excess anisotropy in the ratio of 1.7 : 1 
in the circumstellar background along and orthogonal to the plane of the 
putative disk. They concluded that, if such a disk does exist, its
surface brightness in the near-infrared (and thus its dust mass) would have
to be an order of magnitude lower than the Trilling et al. estimate. 

Our new limit on the dust mass in the 55 Cnc planetary system is a factor
of $\sim$7 lower than that reported by Jayawardhana et al. (2000), which
was already a factor of $\sim$100 below the value reported by Trilling 
\& Brown (1998) based on near-infrared scattered-light observations. We have
now established that our previous sub-millimeter ``detection''
was not exactly coincident with 55 Cnc, but close to one of the peaks 
(source 1) in Figure 1.
The pointing center used by Jayawardhana et al. (2000) was 
fortuitously close to this source due to a coordinate transcription
error.  Given the large ISO beam size of 45$\arcsec\times$45$\arcsec$ per
detector element, it is plausible that 
the Dominik et al. (1998) ISO detection at 60 $\mu$m also originates from 
one or more of these nearby, likely unrelated sources,
and is not due to a dust disk around 55 Cnc. 

In light of our results, those of Schneider et al. (2001), and the possibility
of an unrelated source accounting for the ISO detection, there is little
evidence for a  dust disk around 55 Cnc other than the scattered-light
coronographic images of Trilling \& Brown (1998).
The limits on the amount of small dust particles in the
55 Cnc planetary system from sub-millimeter and HST/NICMOS measurements
are inconsistent with the Trilling \& Brown (1998) estimate by a factor of 10 at best
and 700 at worst. As Schneider et al. suggests, the most likely explanation
for this significant discrepancy is that the detection reported by Trilling
\& Brown (1998) is spurious. 

\section{Summary}
We have obtained sensitive sub-millimeter observations towards
(and around) the 55 Cnc planetary system. We detect three peaks at the
2.5 mJy level in our 850 $\mu$m map that are offset from 
55 Cnc by 25$-$40$\arcsec$. Optical coronagraphic observations show
three point sources over the same region, with two sources north of
55 Cnc approximately aligned with two 850 $\mu$m peaks.
These peaks are likely unrelated background sources.  Our results, together
with the HST/NICMOS observations of Schneider et al. (2001), place 
stringent limits on the amount of dust in the 55 Cnc system, and are
inconsistent with the existence of a circumstellar dust disk as
reported by Dominik et al. (1998) and Trilling \& Brown (1998).  

\bigskip

$\bf Acknowledgements:$ The observations reported here were obtained with the 
James Clerk Maxwell 
Telescope, which is operated by the Joint Astronomy Centre on behalf of the 
Particle Physics and Astronomy Research Council of the United Kingdom, the 
Netherlands Organisation for Scientific Research, and the National Research 
Council of Canada. We wish to thank the JCMT staff for their outstanding
support. Data from the University of Hawaii 2.2-m telescope were
obtained during an observing campaign partly supported by a NASA Origins grant
to David C. Jewitt.  PK was supported by the NSF Center for Adaptive Optics,
managed by the University of California, Santa Cruz, under cooperative
agreement AST 98-76783.


\newpage
\begin{figure}
\plotone{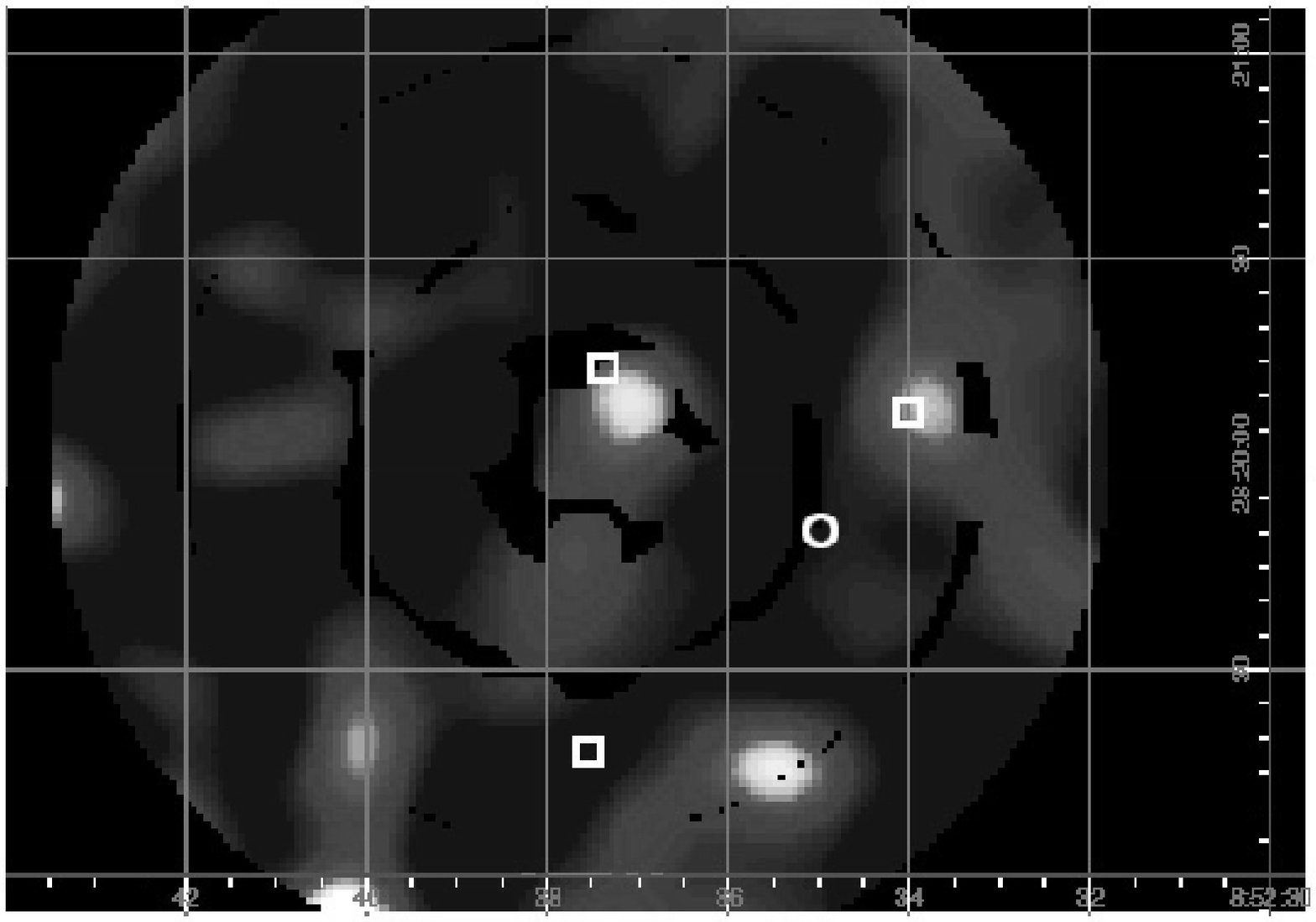}
\caption{JCMT/SCUBA 850 $\mu$m map of the region surrounding 55 Cnc. The 
star's position is marked by the white circle. Positions of optical $R$-band
sources are marked by white boxes.  The map is produced from photometry
observations at two different pointings. Each individual pointing produces 
a map with concentric rings.  Even with two separate pointings, combined with 
the effects of sky rotation, a fully-sampled map is not produced and hence 
residual holes are evident.
}
\end{figure}

\newpage
\begin{figure}
\plotone{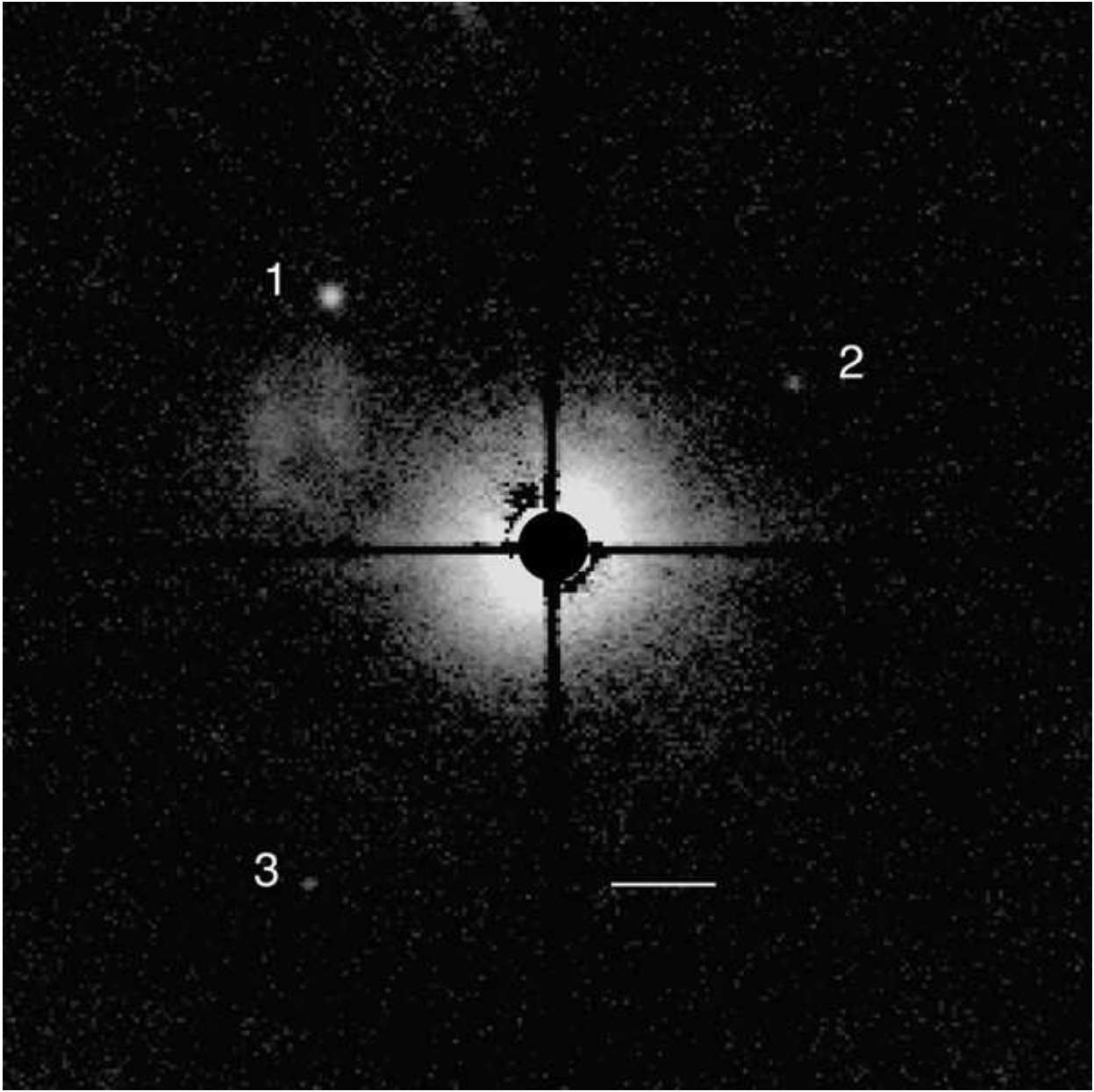}
\caption{R-band coronagraphic image of 55 Cnc after PSF subtraction.  North 
is up, East is left, and the bar represents 10$\arcsec$.  A 6.5$\arcsec$ 
diameter, hard-edged occulting spot is supported by four wires at the 
telescope focal plane.  Spurious instrumental scattered light is evident 
as an extended feature northeast of 55 Cnc and south of Source 1. }
\end{figure}

\clearpage
\begin{table}
\begin{scriptsize}
\begin{center}
\renewcommand{\arraystretch}{1.2}
\begin{tabular}{lcccccccc}
\multicolumn{9}{c}{\scriptsize TABLE 1}\\
\multicolumn{9}{c}{\scriptsize }\\
\multicolumn{9}{c}{\scriptsize POSITIONS AND FLUXES OF SUB-MM AND OPTICAL SOURCES}\\
\multicolumn{9}{c}{\scriptsize }\\
\hline
 & \multicolumn{3}{c}{\scriptsize Sub-mm} & \multicolumn{3}{c}{\scriptsize Optical} & \multicolumn{2}{c}{\scriptsize Difference (arcsec)}\\
\hline
     & RA (2000) & Dec (2000) & 850 $\mu$m flux & RA (2000) & Dec (2000) & $R$ mag. & RA & Dec\\
\hline
Source 1 & 08:52:37.1 & 28:20:09.0 & 2.7 mJy & 08 52 37.4 & +28 20 14.6 & 19.0 & 4.0  & 5.6\\
Source 2 & 08:52:33.8 & 28:20:09.0  & 2.4 mJy & 08 52 34.0 & +28 20 06.3 & 21.1 & 4.0  & 2.7\\ 
Source 3 & 08:52:35.4 & 28:19:16.0 & 2.8 mJy & 08 52 37.5 & +28 19 18.2 & 21.5 & 29.1 & 2.4\\
\hline
\end{tabular}
\end{center}
\end{scriptsize}
\end{table}

\end{document}